\documentclass[preprint,12pt]{aastex}     
\usepackage{graphicx}
 
\def\cc{\,{\rm cm^{-3}}}
\def\cm2{\,{\rm cm^{-2}}}
\def\kms{\,{\rm {km\,s^{-1}}}}

\def\h2{\,{\rm H_{2}}}
\def\13co{\,{\rm ^{13}CO}}
\def\aua{{\rm A\&A} }
\def\auas{{\rm A\&AS} }
\def\apj{{\rm ApJ} }

\def\apjs{{\rm ApJS} }

\newcommand{\co}{\mbox{\rm $^{12}$CO}}
\newcommand{\cothree}{\mbox{\rm $^{13}$CO}}

\newcommand{\hii}{\mbox{\rm \ion{H}{2}}}

\newcommand{\jone}{($J=1\rightarrow0$)}
\newcommand{\jtwo}{($J=2\rightarrow1$)}

\newcommand{\jfour}{($J=4\rightarrow3$)}

\newcommand{\Kkmpers}{K~km~s$^{-1}$}

\begin{document} 
 
\title{High Excitation Molecular Gas in the Magellanic Clouds}
 
 
\author{Alberto D. Bolatto}
\affil{Department of Astronomy \& Radio Astronomy Laboratory}
\affil{University of California at Berkeley, \\
601 Campbell Hall, Berkeley, CA 94720-3411, USA}
\email{bolatto@astro.berkeley.edu}
\author{Frank P. Israel}
\affil{Sterrewacht Leiden, Leiden University} 
\affil{P.O. Box 9513, NL 2300-RA Leiden, The Netherlands}
\email{israel@strw.leidenuniv.nl}
\and
\author{Christopher L. Martin} 
\affil{Department of Physics and Astronomy}
\affil{Oberlin College, Wright Laboratory of Physics, \\
110 North Professor Street, Oberlin, Ohio 44074-1088 USA}
\email{Chris.Martin@oberlin.edu}
 
\begin{abstract} 
We present the first survey of submillimeter \co\ \jfour\ emission in
the Magellanic Clouds. The survey is comprised of 15
$6\arcmin\times6\arcmin$ maps obtained using the AST/RO telescope
toward the molecular peaks of the Large and Small Magellanic
Clouds. We have used these data to constrain the physical conditions
in these objects, in particular their molecular gas density and
temperature. We find that there are significant amounts of molecular
gas associated with most of these molecular peaks, and that high
molecular gas temperatures are pervasive throughout our sample. We
discuss whether this may be due to the low metallicities and the
associated dearth of gas coolants in the Clouds, and conclude that the
present sample is insufficient to assert this effect. 
\end{abstract}
 
\shortauthors{Bolatto, Israel, \& Martin}
\shorttitle{High Excitation Gas in the Magellanic Clouds}


\section{Introduction} 

With their proximity, their unextinguished lines of sight, and their
profuse star formation, the Magellanic Clouds are some of the best
extragalactic objects in which to study the relationship between
molecules and star formation. Because the interstellar medium (ISM) in
the Magellanic Clouds is deficient in heavy elements and dust,
molecular observations of these objects probe an interesting regime,
perhaps more similar to the conditions in early protogalaxies rather
than those prevalent today in the Milky Way. Indeed, studies of
active, metal--poor, nearby dwarf galaxies such as the Magellanic
Clouds should offer insight into the processes at work in primeval
sources.

\begin{figure}
\includegraphics{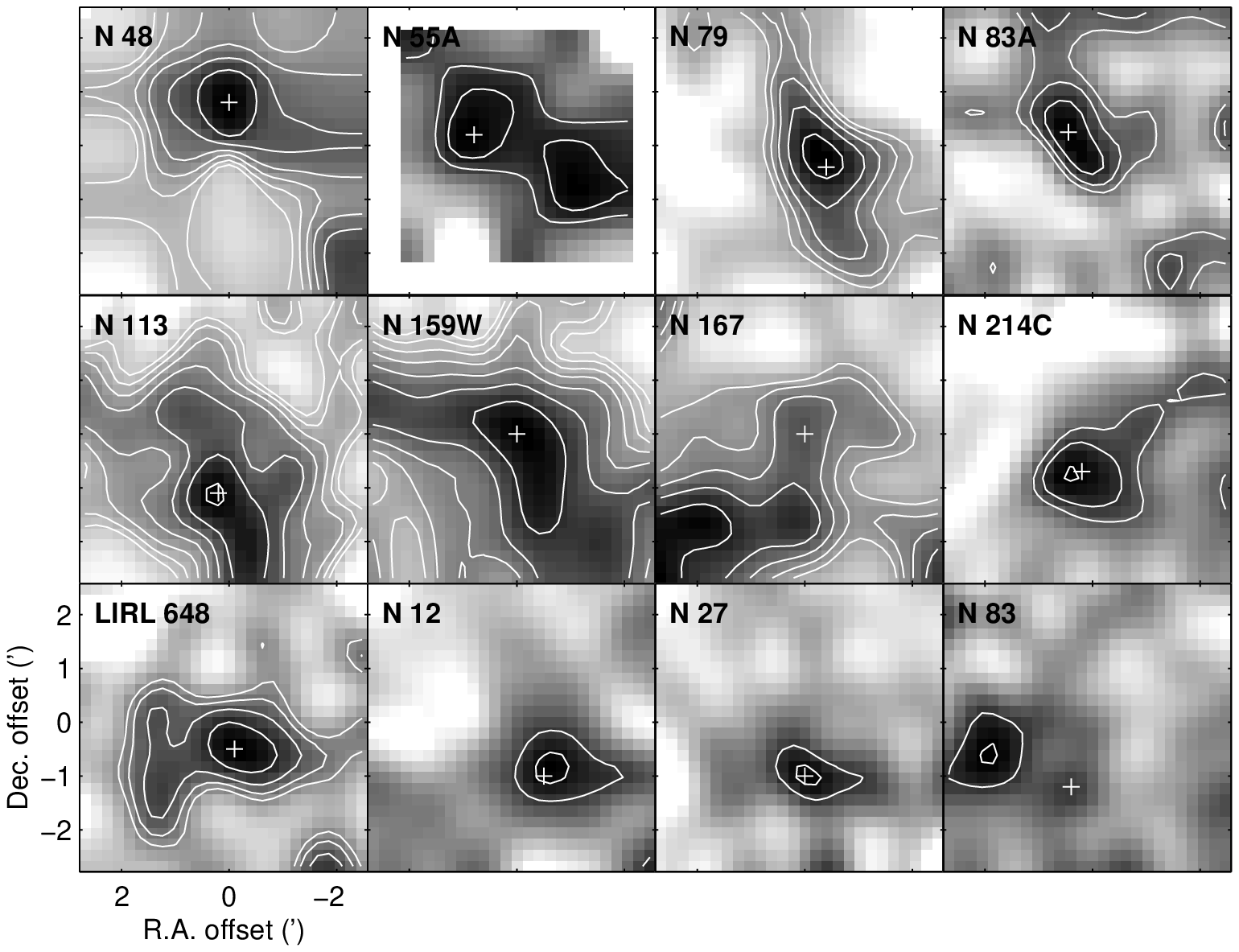}
\caption{\co\ \jfour\ maps obtained toward our detected sources in the
LMC/SMC.  The coordinates of the central positions of the maps
(offsets 0,0) are listed in Table 1. The contours are logarithmically
spaced, starting at 1.5 \Kkmpers\ and increasing by factors of 1.3
(except for N~79, N~83A, N~113, and N~159W, where they start at 2.5
\Kkmpers). These maps were produced by convolving the data with a
Gaussian of FWHM=1\arcmin. Most maps are sampled on a 30\arcsec\ grid,
except for N~48 and N~55A where the grid spacing used was 60\arcsec.
The positions marked by the crosses are those of the CO \jone\
sources, where we measured the intensities used in the LVG analysis.
\label{maps}}
\end{figure}

What are the effects of low metallicities on the star--forming
molecular ISM? We know that molecules are more difficult to form both
in the gas phase (less O, C, and N) and on grain surfaces (fewer
grains), and that they are easier to destroy (diminished dust
shielding of the UV radiation) in low metallicity environments. Thus,
molecules other than $\h2$ are more rare in these sources which in
particular translates into a dearth of CO emission in the Magellanic
Clouds and other dwarf irregulars (e.g., Israel et al., 1986; Israel,
Tacconi \& Baas 1995, Taylor, Kobulnicky, \& Skillman 1998; Leroy et
al. 2005). Furthermore, because the FIR and submillimeter lines of the
different forms of carbon and oxygen (C$^+$, C, O, and CO) dominate
the cooling of the star--forming ISM (e.g., Le Bourlot et al. 1993;
Wolfire et al. 1995), the lower abundances of these elements will make
the cooling of the molecular gas less efficient. However, at the same
time smaller dust--to--gas ratios will yield lower heating of the
molecular gas, as photoelectric ejection of electrons from small dust
grains is the chief mode by which starlight heats the gas phase of the
ISM. If molecular gas temperatures were considerably affected by the
metallicity of the ISM, we expect important consequences for star
formation in such environments. In particular, if the Jeans criterion
is relevant to star--formation, the mass of collapsable clouds grows
for decreasing metallicity as the Jeans mass increases $M_J\sim
T^{3/2}$. Such change could have important effects on the Initial Mass
Function of stars in these systems.  To a first approximation, models
suggest that lowering the metallicity causes a similar decrement in
both heating (by diminishing the dust--to--gas ratio) and cooling (by
diminishing the C and O abundances; Wolfire et al. 1995).

It is important to realize, however, that there are many possibilities
likely to complicate this simple picture. For example, if the
dust--to--gas ratio were to decrease faster than the metallicity (as
suggested by Lisenfeld \& Ferrara 1998), or if there were a lack of
very small dust grains in the low metallicity ISM (as suggested by the
faintness of the polycyclic aromatic hydrocarbon emission observed
toward some of these sources; e.g., Madden 2000), the heating
processes may become less efficient and the balance may be shifted
toward lower temperatures. Furthermore, because it is also necessary
to consider the metallicity threshold below which hitherto secondary
heat sources (e.g., chemical heating) become important, the effects of
metallicity on the heating and cooling balance of molecular clouds are
very difficult to address from a purely theoretical approach.

Answering some of these questions observationally requires studying
the physical conditions of the molecular gas in nearby, low
metallicity sources.  At distances of 55 and 63 kpc, the proximity of
the Magellanic Clouds affords single--dish millimeter--wave
observations excellent spatial resolution attainable in other galaxies
only through the use of interferometers, which permits detailed
studies of individual clouds instead of ensemble properties. In
particular, the ability to spatially separate the emission from
different regions makes the Magellanic Clouds ideal targets to study
the excitation of the molecular gas and its relationship with star
formation. Multitransition studies of CO and other molecules are very
useful tools to determine the physical conditions of the H$_2$ (e.g.,
Johansson et al. 1998; Heikkil\"a 1998; Chin et al. 1998; Heikkil\"a
et al. 1999), but their application is limited if there is no
information on the intensities of the higher CO transitions, which are
extremely sensitive to density and temperature.  Because of their
southern declination, however, there is a dearth of submillimeter
observations of the Clouds.

\begin{figure}
\plotone{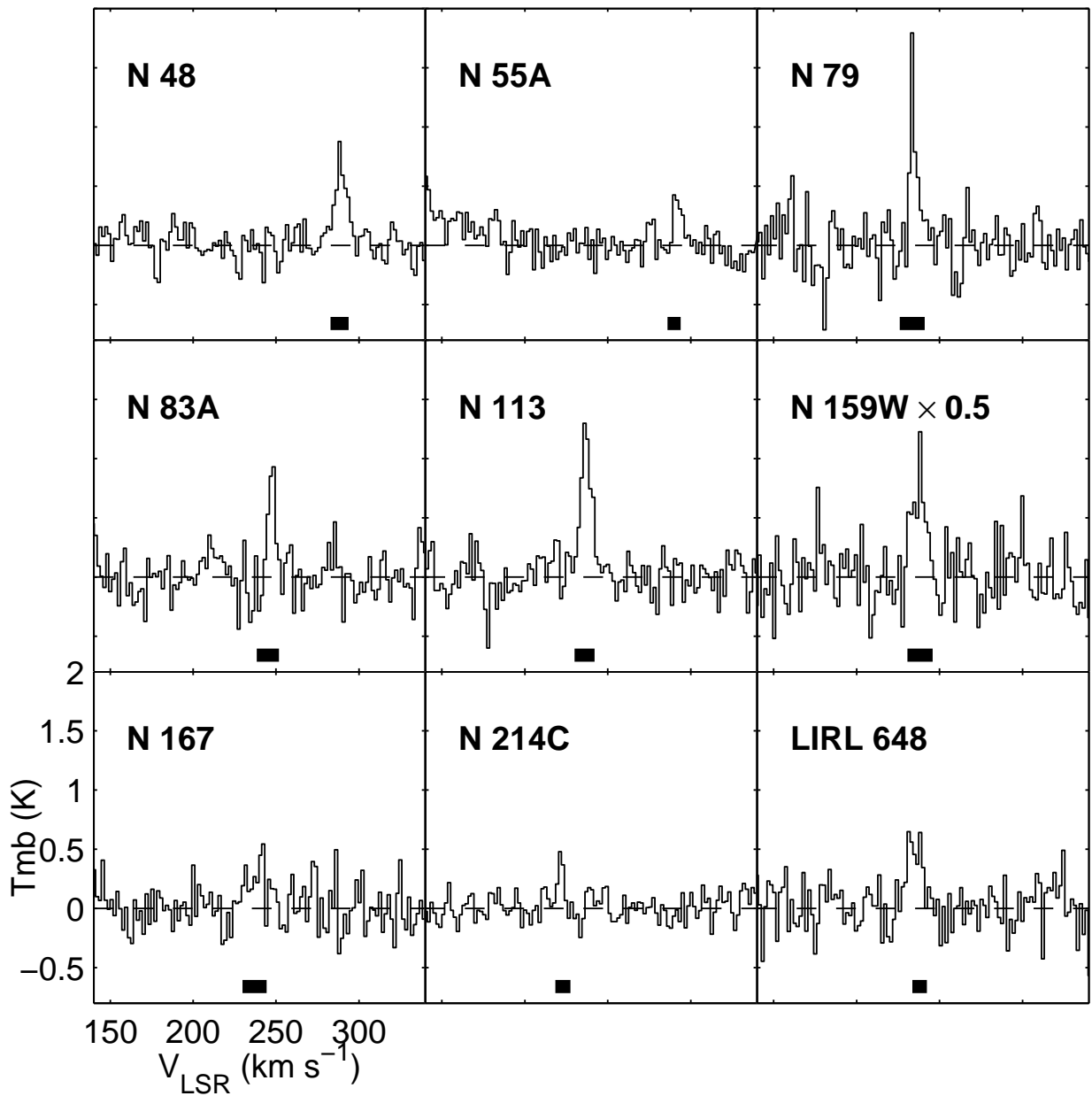}
\caption{Spectra of \co\ \jfour\ emission in LMC sources.  These
spectra were obtained toward the positions marked by crosses in
Fig. \protect\ref{maps}. All sources were detected. The emission in
N~159W is shown scaled down by a factor of two for display purposes. The
thick black line near the bottom of the plots indicates the velocity
and width of the CO emission in the lower $J$ transitions, as observed
using SEST.\label{lmcspec}}
\end{figure}

We present here a survey of \co\ \jfour\ in the molecular peaks of the
Magellanic Clouds, and the results of an excitation analysis using
these and lower $J$ observations. These pointings were selected among
the brightest \co\ \jone\ peaks found by SEST observations (Israel et
al. 1993), many of which are associated with star--forming complexes
and Henize (1956) H$\alpha$ nebulosities, and are thus denoted using
the corresponding ``N'' number. In section \S\ref{obs} we present the
observations, in \S\ref{disc} we discuss the LVG analysis and its
results, and in \S\ref{summ} we summarize our conclusions.

\section{Observations and Results}
\label{obs}

\begin{center}
\begin{figure}
\plotone{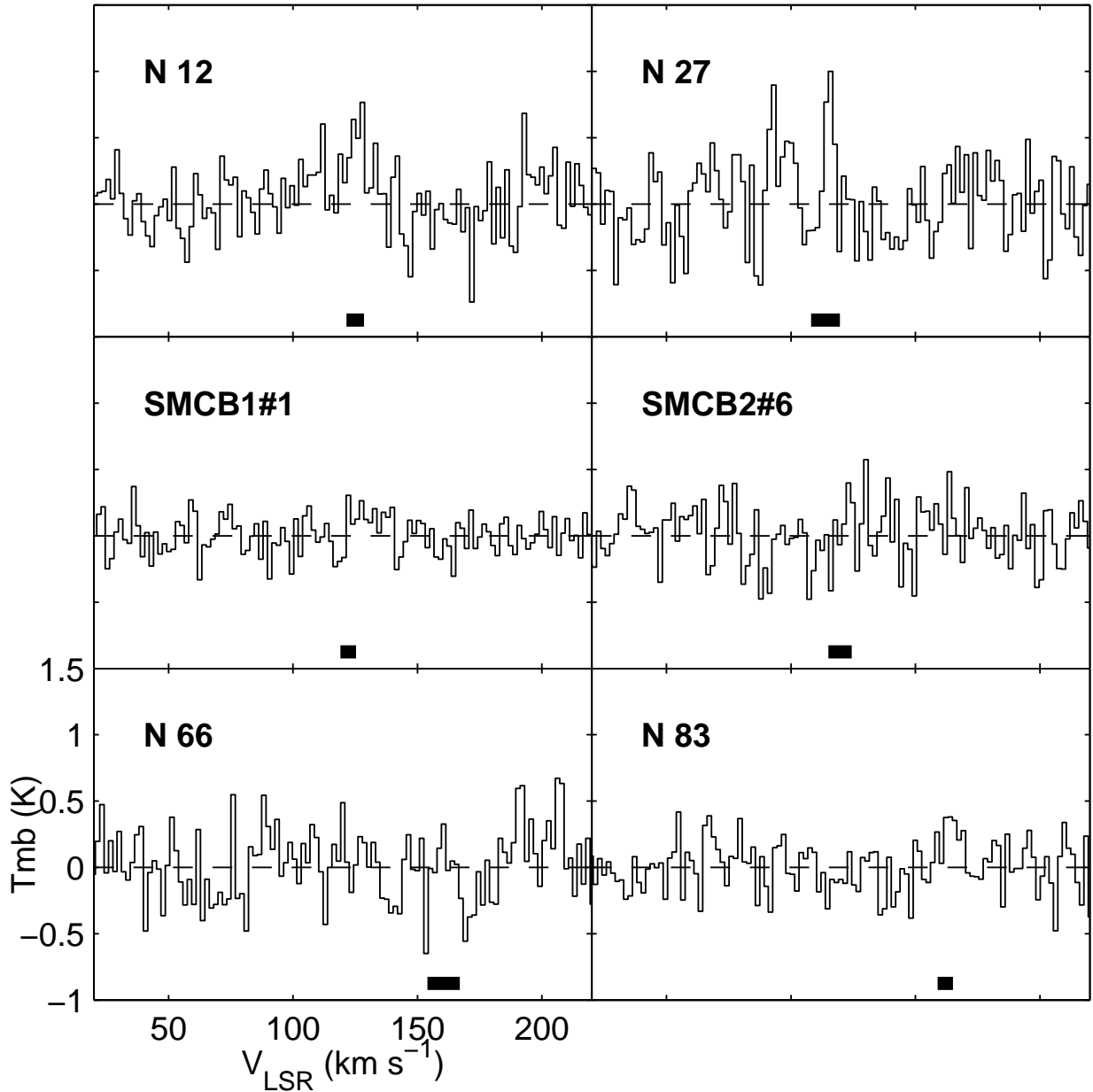}
\caption{Spectra of \co\ \jfour\ emission in SMC sources, obtained
toward the positions marked with crosses in
Fig. \protect\ref{maps}. Only N~12, N~27, and N~83 were detected, and the
signals are in general considerably weaker than in the LMC. As in the
previous figure, the CO \jone\ velocity and linewidth are indicated by
the thick black line at the bottom of the plots.\label{smcspec}}
\end{figure}
\end{center}

We observed the \jfour\ transition of carbon monoxide (\co) at
$\nu\simeq 461.0408$ GHz (650.7 $\mu$m) using AST/RO, the Antarctic
Submillimeter Telescope and Remote Observatory located at the
Amundsen--Scott South Pole base (Stark et al. 2001).  The observations
were obtained on the austral winter of 2002, using the lower frequency
side of the dual AST/RO SIS waveguide receiver (Walker et al. 1992;
Honingh et al. 1997), with system temperatures of
$T_{sys}\sim2000$~K. The backend was the 2048 channel low resolution
(1.07 MHz resolution, 0.68 MHz channels) acousto--optical spectrometer
(Schieder, Tolls, and Winnewisser 1989). The spectra were observed in
position switching mode, chopping 25\arcmin\ in Azimuth (which is the
same as R.A. at the pole). At 461 GHz, the telescope beam was measured
to have a HPBW$\approx109\arcsec$. The forward efficiency determined
from skydips was $\sim70\%$, and is assumed to be identical to
$\eta_{mb}$. Maps of size $6\arcmin\times6\arcmin$ centered on the
SEST coordinates (c.f. Table 1) were obtained for each region, using a
30\arcsec\ grid (in a few cases the maps were done on a 60\arcsec\
grid).  The data were calibrated using the standard procedure for
AST/RO, which includes sky, ambient and cold load measurements every
20--30 minutes, and processed using the COMB astronomical package.
The individual maps are shown in Fig. \ref{maps}, and representative
spectra are shown in Figs. \ref{lmcspec} and \ref{smcspec}. Because
the pointing accuracy is estimated to be $\sim1\arcmin$, we have
selected the emission peaks closest to the center of the map within
those margins to measure the integrated intensities compiled in Table
1 (indicated by the crosses in Fig. \ref{maps}). Note that this method
may introduce a bias in the direction of obtaining larger CO
\jfour/\jone\ ratios. Nonetheless we feel that this methodology is
justified, as in the Milky Way the positional coincidence between the
peaks of both transitions is very often observed. In some cases it is
apparent that the structure of the sources is complex (e.g., N~159W;
Bolatto et al. 2000), and in particular for N~167 and N~83 the
brightest emission peak is well away from the central position. Given
the large and variable effects of the atmosphere at submillimeter
wavelengths and the pointing accuracy of the telescope, we estimate
the overall absolute calibration accuracy of the data to be
$\sim30\%$.

Table 1 summarizes our observations. Column 4 lists the \co\ \jone\
integrated intensities observed by the SEST Magellanic Cloud Key
Programme toward our sources, while column 5 shows the intensities
convolved to the angular resolution of AST/RO. Finally, column 6 lists
the \co\ \jfour\ integrated intensities used in our LVG analysis, with
their statistical errors. These intensities are those measured in the
AST/RO pointings marked with crosses in Fig. \ref{maps}.

\begin{deluxetable}{lrrrrr}
\tabletypesize{\small}
\tablewidth{0pt}
\tablecaption{AST/RO and SEST CO observations of Magellanic cloud \hii\ regions}
\tablehead{
Source & \multicolumn{2}{c}{Coordinates} & \multicolumn{3}{c}{Observed \co\ Integrated Intensities in \Kkmpers}      \\
	& \multicolumn{1}{c}{R.A.}   & \multicolumn{1}{c}{Dec.}  & \multicolumn{1}{c}{($43''$)}  & \multicolumn{1}{c}{($109''$)} & \multicolumn{1}{c}{($109''$)}         \\
        &  \multicolumn{1}{c}{(B1950)} & \multicolumn{1}{c}{(B1950)} & \multicolumn{1}{c}{\jone}  & \multicolumn{1}{c}{\jone}   &  \multicolumn{1}{c}{\jfour}          \\
}            
\startdata
\cutinhead{\bf LMC}                 \\
N~48	&  $05^{\rm h}25^{\rm m}46\fs6$ & $-66^\circ17\arcmin36\arcsec$ & 3.7       &   \nodata     & 6.6$\pm$0.7 \\
N~55A	&  $05^{\rm h}32^{\rm m}30\fs0$ & $-66^\circ29\arcmin21\arcsec$ & 17.4      &   4.9     & 2.5$\pm$0.5 \\
N~79     &  $04^{\rm h}52^{\rm m}09\fs5$ & $-69^\circ28\arcmin21\arcsec$ & 21.3      &   \nodata     & 6.0$\pm$0.7 \\
N~83A	&  $04^{\rm h}54^{\rm m}17\fs0$ & $-69^\circ16\arcmin23\arcsec$ & 21.1      &   6.6     & 10.4$\pm$1.4\\
N~113	&  $05^{\rm h}13^{\rm m}40\fs2$ & $-69^\circ25\arcmin37\arcsec$ & 23.4      &   \nodata     & 9.7$\pm$0.9 \\
N~159W   &  $05^{\rm h}40^{\rm m}01\fs5$ & $-69^\circ47\arcmin02\arcsec$ & 57.0      &  45.6     & 20.8$\pm$2.2\\
N~167    &  $05^{\rm h}44^{\rm m}17\fs6$ & $-69^\circ23\arcmin19\arcsec$ & 19.8      &  15.5     & 5.6$\pm$1.1 \\
N~214C	&  $05^{\rm h}42^{\rm m}21\fs8$ & $-71^\circ20\arcmin33\arcsec$ &  9.2      &   \nodata     & 2.0$\pm$0.4 \\	
LIRL~648	&  $05^{\rm h}14^{\rm m}07\fs0$ & $-69^\circ38\arcmin57\arcsec$ & 13.1      &   \nodata     & 6.7$\pm$1.2 \\
\cutinhead{\bf SMC}\\
N~12	&  $00^{\rm h}44^{\rm m}50\fs5$ & $-73^\circ22\arcmin33\arcsec$ &  9.0      &  3.46     & 3.85$\pm$0.7      \\
N~27	&  $00^{\rm h}46^{\rm m}32\fs9$ & $-73^\circ21\arcmin50\arcsec$ & 11.6      &  5.63     & 5.0$\pm$1.3       \\  
SMCB1\#1	&  $00^{\rm h}43^{\rm m}42\fs0$ & $-73^\circ35\arcmin10\arcsec$ &  4.1      &  2.67     & $\leq$ 0.5        \\
SMCB2\#6	&  $00^{\rm h}46^{\rm m}28\fs1$ & $-73^\circ34\arcmin10\arcsec$ &  6.4      &  3.20     & 1.7$\pm$0.9       \\
N~66	&  $00^{\rm h}57^{\rm m}26\fs5$ & $-72^\circ26\arcmin36\arcsec$ &  0.9      &  3.41\tablenotemark{a}      & 1.7$\pm$1.1       \\
N~83	&  $01^{\rm h}12^{\rm m}29\fs2$ & $-73^\circ32\arcmin40\arcsec$ &  5.6      &  2.66     & 2.0$\pm$0.4       \\
\enddata
\label{intensities}
\tablenotetext{a}{Obtained from the \cothree\ \jone\ map convolved to 
a HPBW=109\arcsec, assuming a \co/\cothree\ ratio $\approx11$.}
\tablecomments{Nominal map center coordinates are indicated. Column 5 
contains the \co\ \jone\ integrated intensity in the \jfour\ beam
for sources with \jone\ maps.}
\end{deluxetable}

\begin{center}
\begin{deluxetable}{lcccccl}
\tabletypesize{\footnotesize}
\tablewidth{0pt}
\tablecaption{CO line intensity ratios of Magellanic Cloud Objects}
\tablehead{
Source	& \multicolumn{3}{c}{\co\ Transition Ratios} & \multicolumn{2}{c}{\co/\cothree\ Isotopic Ratios} & References \\
        & ($2\rightarrow1/1\rightarrow0$)   & ($3\rightarrow2/1\rightarrow0$) & ($4\rightarrow3/1\rightarrow0$)   & ($1\rightarrow0$) & ($2\rightarrow1$)  & }
\startdata
\cutinhead{\bf LMC}\\
N~55A	& \nodata           & \nodata         & 0.51$\pm$0.17 &  11   &  \nodata   & 1, 2\\
N~83A	& \nodata           & \nodata         & 1.58$\pm$0.52 &   9   &  \nodata   & 1, 3\\
N~159W   & 0.90$\pm$0.25 & 0.9         & 0.46$\pm$0.15 &   9   &   5    & 1, 4, 5\\
N167    & 1.16$\pm$0.23 & \nodata         & 0.36$\pm$0.12 &  12   &  \nodata   & 1, 6\\
\cutinhead{\bf SMC}\\
N~12	& 1.20$\pm$0.40 & 1.0        & 1.11$\pm$0.28  &  11   &    9   & 1, 7\\
N~27	& 0.95$\pm$0.35 & 0.75       & 0.89$\pm$0.30  &  17   &   11   & 1, 8, 9\\  
SMCB1\#1	& 0.72$\pm$0.12 & 0.4        & $\leq$0.2      &  11   &  13.5  & 1, 10, 11\\
SMCB2\#6	&  \nodata          &  \nodata       & 0.53$\pm$0.28  &  \nodata  &   \nodata  & 1   \\
N~66	& 1.30$\pm$0.35 & 1.0        & 0.50$\pm$0.32  &  11   &    7   & 1, 9, 10\\
N~83	& 1.19$\pm$0.31 &  \nodata       & 0.56$\pm$0.22  &  10   &   9.5  & 1, 12 \\
\enddata
\label{ratios}
\tablecomments{Ratios are determined from values in Table 1, and from previously
published papers, mostly in the ESO Key Programme. References: 
(1) This Paper; (2) Israel et al. (1993); (3) Israel et al. 
(2003); (4) Johansson et al. (1998); (5) Bolatto et al. (2000); (6)
Garay et al. (2002); (7) Chin et al. (1998); (8) Heikkil\"a et al. (1999); 
(9) Rubio et al. (1996); (10) Heikkil\"a (1998); (11) Rubio et al. (1993); 
(12) Bolatto et al. (2003).}
\end{deluxetable}
\end{center}

\section{Analysis and Discussion}
\label{disc}

\subsection{Modeling of CO}

\begin{center}
\begin{deluxetable}{lccc}
\tabletypesize{\small}
\tablewidth{0pt}
\tablecaption{Model physical parameters -- single component fits}
\tablehead{
Source	& Kinetic     & Volume        & Column     \\
        & Temperature & Density       & Density    \\
        & $T_{\rm k}$ & $n(\h2)$      & $N$(CO)/dV \\ 
        & (K)         & ($\cc$)       & ($\cm2\hbox{km$^{-1}$ s}$)
}
\startdata
\cutinhead{\bf LMC} \\
N~55    &  30--60     &3$\times10^{3}$&   $1\times10^{17}$        \\
N~79    &  100        &5$\times10^{3}$&  6-$10\times10^{17}$      \\
N~113   &  150        &3$\times10^{3}$&   $6\times10^{17}$        \\
N~159W  &  150        & 10$^{3}$      &   $3\times10^{17}$        \\
LIRL~648 &   60        & 10$^{4}$      &   $3\times10^{17}$        \\
\cutinhead{\bf SMC} \\
SMCB1\#1  &   50        & 7$\times10^{2}$      &  $0.5\times10^{17}$       \\
\enddata
\label{singlefit}
\end{deluxetable}
\end{center}

The available observed \co\ and \cothree\ line ratios have been
modelled using the large--velocity gradient (LVG) radiative transfer
models described by Jansen (1995) and Jansen et al. (1994). These
models provide line intensities as a function of three input
parameters: gas kinetic temperature $T_{\rm k}$, molecular hydrogen
density $n(H_{2})$ and CO column density per unit velocity ($N({\rm
CO})$/d$V$). By comparing model line {\it ratios} to the observed
ratios we determine the physical parameters best describing the
conditions in the observed source. In principle, with two isotopes we
need to measure five independent line intensities in order to fully
determine the conditions of a single molecular gas component
(i.e. $T_{\rm k}$, $n(H_{2})$, $N(\co)$/d$V$, $N(\cothree)$/d$V$ and a
beam filling--factor). By assuming a fixed isotopical abundance
[\co]/[\cothree] = 40 (Johansson et al.  1994) we may decrease this
requirement to four independent line intensities. As Table 2 shows,
this minimum requirement is met by two out of nine LMC objects and
five out of six SMC objects. The physical conditions of the remaining
eight objects are, in principle, underdetermined.

More realistic, and consequently more complex, models of gas
excitation that include more than one component require many more
observations to be properly constrained.  Full modelling of a
two--component molecular cloud using two isotopes requires ten
independent measurements, which again are reduced to eight by the
introduction of a fixed isotopical abundance. As this is more than we
have actually observed in any of the LMC or SMC clouds, it is clear
that the solutions may not be unique.  As long as the range of 
possible solutions is not excessive, however, they are still useful to
constrain the physical parameters governing the observed
emission.

We identified acceptable fits by searching a grid of model parameter
combinations (10 K $\leq T_{\rm k} \leq $ 150 K, $10^{2} \cc \leq
n(\h2) \leq 10^{5} \cc$, $6 \times 10^{15}\,\cm2\,(\kms)^{-1} \leq
N($CO$)/$d$V \leq \,3 \times 10^{18}\,\cm2\,(\kms)^{-1}$) for model
line ratios matching the observed values. Although errors in the line
ratios increase the range of possible solutions, these ratios tend to
define reasonably well--constrained regions of parameter space. The
solutions are somewhat degenerate, as variations in the parameters may
compensate one another. For instance, a simultaneous increase in
kinetic temperature and decrease in $\h2$ densities (or vice versa)
yields similar line ratios (see \S\,3.2.3).

\subsection{Single--component fits}

\subsubsection{Objects with two measured intensities}

For three objects (LMC-N~48, LMC-N~214C, and SMCB2\#6) we only have
intensities in the \jone\ and \jfour\ transitions of \co.  The
parameters of these clouds are thus poorly constrained, and not
summarized in a table.  Assuming a single molecular gas component, we
find for SMCB2\#6 no effective constraints: $T_{\rm k}$ = 20--150 K,
$n(\h2) = 10^{2}-10^{4}\,\cc$, and $N($CO$)/$d$V =
10^{16}-10^{18}\,\cm2 (\kms)^{-1}$.  All \co\ transitions should be
very optically thick. LMC-N~214C is slightly better determined:
temperatures below 30 K are not allowed and densities appear high
$n(\h2) = 10^{4}-10^{5}\,\cc$. The \jone\ transition should not be
very optically thick (although the higher transitions are) and its
isotopic intensity ratio should be 10 -- 20, as is in fact commonly
observed in the LMC (Israel et al. 2003). Finally, the high \co\
\jfour/\jone\ ratio exhibited by LMC-N~48 are, in the
single--component approximation, only consistent with low optical
depths (isotopic intensity ratios 20 -- 40), rather high densities
$n(\h2) = 10^{4}-10^{5}\,\cc$ and temperatures $T_{\rm k}$ = 60--150
K, and gradients $N($CO$)/$d$V = 10^{16}-10^{17}\,\cm2\,(\kms)^{-1}$.

\subsubsection{Objects with three measured intensities}

For five objects, all in the LMC, (N~55, N~79, N~83A, N~113 and
LIRL~648) we have measured \cothree\ \jone\ intensities in addition to
the \co\ \jone\ and \jfour\ intensities. As is clear from the previous
discussion, the molecular gas parameters are still underdetermined,
but not fully unconstrained (Table 3). Both N~79 and N~113 fit very well to a
hot and fairly dense model cloud with $T_{\rm k}$ = 100--150 K,
$n(\h2) = 3000-5000\,\cc$ and a gradient of about $6 \times
10^{17}\,\cm2\,(\kms)^{-1}$. LIRL~648 is fit, but not very well, by
a gas at the somewhat lower temperature of 60 K and the somewhat
higher density of $10^4\,\cc$. The physical parameters of the
molecular gas cloud associated with N~83A are inconsistent with the
assumption of a single component. Only a very poor fit is obtained in
the high-temperature, high-density, and high-velocity gradient limit.

\begin{center}
\begin{deluxetable}{lrrccccc}
\tablecaption{Model fits for SMCB1\#1\label{FineSol}}
\tabletypesize{\footnotesize}
\tablewidth{0pt}
\tablehead{
\colhead{$T_{\rm kin}$} & \colhead{$n(\h2)$} & \colhead{$N($CO$)/$d$V$} & \multicolumn{2}{c}{$I\co/I\13co$} &
\multicolumn{3}{c}{$\co$ Rotational Ratio} \\
  \colhead{(K)}  & \colhead{($\cc$)}  & \colhead{($\cm2\,(\kms)^{-1}$)} & \colhead{\jone} & \colhead{\jtwo} & \colhead{(1--0/2--1)} & \colhead{(3--2/2--1)} & \colhead{(4--3/2--1)} 
}
\startdata
\tableline
\multicolumn{3}{l}{\bf Observed constraints}\\
\tableline
\nodata & \nodata &\nodata & 11.2 & 13.5 & 1.4  & 0.6  & $\leq$0.24 \\
\tableline
\multicolumn{3}{l}{\bf Modeling results}\\
\tableline
 60 &  500 &  $6\times10^{16}$   & 10.4 & 12.0 & 1.41 & 0.56 & 0.23 \\
 50 &  600 &  $5\times10^{16}$   & 11.4 & 12.2 & 1.45 & 0.54 & 0.21 \\
{\it50}&{\it700}&${\it5\times10^{16}}$&{\it12.0}&{\it12.2}&{\it1.43}&{\it0.56}&{\it0.23}\\
 50 &  800 &  $5\times10^{16}$   & 12.4 & 11.8 & 1.40 & 0.57 & 0.24 \\
 40 & 1100 &  $5\times10^{16}$   & 12.4 & 11.2 & 1.37 & 0.58 & 0.24 \\
 30 & 2000 &  $4\times10^{16}$   & 14.1 & 10.8 & 1.30 & 0.59 & 0.24 \\
\enddata
\end{deluxetable}
\end{center}

\begin{center}
\begin{deluxetable}{lccccccc}
\tablecaption{Model physical parameters -- dual component fits}
\tablewidth{0pt}
\tabletypesize{\footnotesize}
\tablehead{
Source	& \multicolumn{3}{c}{`Cold Dense' Component} & \multicolumn{3}{c}{`Hot Tenuous' Component} & Relative\\
        & Kinetic     & Volume       & Column & Kinetic     & Volume       & Column     & $J$=2--1\\
        & Temperature & Density      & Density & Temperature & Density      & Density   & $\co$  \\
        & $T_{\rm k}$ & $n(\h2)$     & $N(CO)/dV$ & $T_{\rm k}$ & $n(\h2)$ & $N(CO)/dV$ & Emission \\ 
        & (K) & ($\cc$) & ($\cm2\hbox{km$^{-1}$ s}$) & (K) & ($\cc)$    & ($\cm2\hbox{km$^{-1}$ s}$) &
}
\startdata
\cutinhead{\bf LMC}
N~83A   &  10  & 10$^{5}$    & $10\times10^{17}$    & 100  & 10$^{5}$  & $3\times10^{17}$    & 3:2 \\ 
        &  60  & 10$^{5}$    &  $1\times10^{17}$    & 150  &5-10$\times10^{2}$& 6-10$\times10^{17}$ & 4:1 \\
        & 150  & 10$^{5}$    &  $3\times10^{17}$    & 150  & 10$^{2}$  & 6-10$\times10^{17}$ & 2:3 \\ 
N~159W  &  20  & 10$^{5}$    &  $1\times10^{17}$    & 100  & 10$^{2}$  & $1\times10^{17}$    & 1:1 \\
N~167   &  20  & 10$^{4}$    & $0.6\times10^{17}$   &30--60&1-10$\times10^{2}$&0.3-10$\times10^{17}$& 2:1 \\
\cutinhead{\bf SMC}
N~12    &  150 & 10$^{5}$    & $10\times10^{17}$    & 150  &1-5$\times10^{2}$ &  $6\times10^{17}$   & 1:1 \\
N~27    &  30  & 10$^{5}$    & $0.6\times10^{17}$   &60-100& 10$^{2}$  & $0.6\times10^{17}$  & 1:1 \\
N~66    &20--60&10$^{4}$-10$^{5}$& $6\times10^{17}$ & 300  & 10$^{4}$  & $0.3\times10^{17}$  & 1:9 \\
N~83    &10--30& 10$^{4}$    & 1-2$\times10^{17}$   & 100  &3$\times10^{3}$& $0.3\times10^{17}$  & 1:8 \\
\enddata
\label{dualfit}
\end{deluxetable}
\end{center}

\subsubsection{Objects with four or more measured intensities}

Seven objects have a sufficiently large number of measured line
intensities to allow a full determination of physical parameters,
assuming they can be properly described by a single molecular gas
component. This appears indeed to be the case for SMCB1\#1, where we
find excellent agreement between the observed ratios and those of a
model gas characterized by $T_{\rm k}$ = 50 K, $n(\h2) =
700\,\cc$, $N($CO$)/$d$V = 5 \times 10^{16}\,\cm2\, (\kms)^{-1}$.  To
fine-tune the solutions, as well as to gain insight into the various
trade-offs, we have run a finer grid of models for this particular
source. We summarize the best solutions for the temperature range
$T_{\rm k}$ = 30--60 K in Table\,\ref{FineSol}. The overall best fit 
is achieved using a kinetic temperature $T_{\rm k}$ = 50 K, and a density
$n(\h2) = 700\,\cc$. At constant temperature, the uncertainty in
density is about 20$\%$. From the Table, it is also clear that we may
allow for a similar temperature uncertainty of 20$\%$ if we
simultaneously increase or decrease the density by 40$\%$.

The good fit between model and observed line ratios is of particular
significance because the solution is overdetermined with six
independent intensity measurements. SMCB1\#1 is a small and isolated
molecular cloud in the SMC Bar, not associated with a star-forming
region (Rubio et al.  1993; Reach, et al. 2000; Rubio et al. 2004),
thus a low density of $\sim700$ $\cc$ is not necessarily surprising,
although the kinetic temperature is higher than we would have expected
for such a cloud in the Milky Way. Table \ref{FineSol}, however,
illustrates the difficulty in pinning down the physical conditions
even in a simple cloud with a simple model.

The measured intensities of N~159-W can also be fit, albeit
somewhat poorly, by a single hot ($T_{\rm k}$ = 150 K) and moderately
dense component ($n(\h2) = 1000 \cc$). However, none of the other five
objects have intensities consistent with a single component.
Typically, a component fitting the observed \co\ line ratios would
fail completely to explain the \co/\cothree\ isotopical ratios,
indicating that in addition to a warm gas component the presence of a
second cooler and dense component should be assumed. The
lack of associated luminous objects probably explains the
exceptionally homogeneous nature of SMCB1\#1, required for such a
well-determined single-component fit to be valid.

\subsection{Dual-component fits}

Since (with the exception of SMCB1\#1 and perhaps N~159-W) {\it none} of the
sources for which a lot of constraints are available allows a good fit
with a single gas component, we suspect that the succesful
single-component fits of the sources in Table\,\ref{singlefit} result
primarily from insufficient information rather than from a simple
physical structure. Attempts at more sophisticated modelling achieve
little of value for most of these sources. The single exception is
LMC-N~83A where only two observed ratios nevertheless conflict with
every single component model tried. A dual component model yields
acceptable but, not surprisingly, poorly constrained solutions (see
Table\,\ref{dualfit}).  For all the other well-observed sources where
single-component model fits failed, we have also constructed
dual-component fits which are likewise listed in
Table\,\ref{dualfit}. This table shows that usually one of the two
components component is reasonably well-determined, whereas the
parameters of the other are generally less tightly constrained.

\section{Summary and Conclusions}

A general result of our survey is the detection of significant \co\
\jfour\ emission in most molecular peaks in the Magellanic Clouds. By
itself this demonstrates the widespread occurrence of significant
amounts of warm molecular gas, as tentatively suggested by Israel et
al. (2003). Application of LVG model calculations show that molecular
gas kinetic temperatures as high as $T_{\rm kin}$ = 100 -- 300~K
frequently occur. Detailed analysis of the objects for which multiple
line ratios are available strongly suggests that the higher
temperatures occur in cloud regions that are not very dense ($n_{\h2}
= 10^{2} - 10^{3} \cc$), and that this gas is generally associated
with colder (typically $T_{\rm kin}$ = 10 -- 60 K) but much denser
($n_{\h2} = 10^{4} - 10^{5} \cc$) molecular gas.  The simplified
analysis possible in cases where fewer line ratios are available
strongly suggests that the situation in those objects is similar.  We
note that recently reported observations of the N~44 complex in the
LMC by Kim et al. (2004) show also the presence of strong
\co\ \jfour\ emission, leading the authors to conclude the likely
presence of very high densities ($n_{\h2} \approx 10^{5} \cc$).

A few clouds are notable in our survey. Already mentioned is the case
of SMCB1\#1, a quiescent, compact molecular cloud devoid of any star
forming activity. It appears to be essentially homogeneous and of
modest density ($n_{\h2} = 700 \cc$), but is surprisingly warm
($T_{\rm kin}$ = 50 K). As there are no known embedded heating
sources, this temperature must be maintained by the environment of the
cloud, located in the south west region of the SMC Bar. The other two
notable objects also occur in the SMC. N~12 is likewise located in the
southern part of the Bar. Here, both the dense and relatively tenuous
molecular phases appear to be surprisingly hot ($T_{\rm kin}$ = 150
K).  In N~66, a very hot ($T_{\rm kin}$ = 300 K) and a cooler ($T_{\rm
kin} = 40\pm20$ K) phase coexist, {\it both} at a rather high density
of the order of $n_{\h2} = 10^{4} \cc$. This particular molecular
cloud is a relatively small remnant in a large and luminous
star-forming complex (c.f., Rubio et al. 1996). It is almost certainly
in an advanced stage of destructive processing; high densities and
temperatures are consistent with such a situation (Rubio et al. 2000;
Contursi et al. 2000).

Have we found signs of metallicity effects on the temperature
equilibrium of the ISM in metal--poor environments? Given the
measurement uncertainties and the dearth of comparable datasets on
``normal'' clouds in our own galaxy such finding cannot be asserted on
the present measurements.  It is suggestive, however, that high
temperatures seem pervasive even in largely quiescent clouds such as
SMCB1\#1. This result should be taken with caution, as we suffer from
a strong sample selection bias: the brightness of the optically thick
\co\ \jone\ transition is proportional to the source temperature and
beam filling fraction, and we have pointed toward the brightest \co\
\jone\ clouds. Thus it may not be surprising that our clouds are
warm. Furthermore, studies of Galactic photodissociation regions
(e.g., S~140, Draine \& Bertoldi 1999; NGC~2023, Draine \& Bertoldi
2000) find that H$_2$ in star--forming clouds is frequently warmer
than the theoretical expectation. Similarly, the intensities of the
mid--J CO lines tend to be underpredicted by homogeneous PDR
calculations (Hollenbach \& Tielens 1999 and references therein).  It
is unclear whether these discrepancies are due to the presence of
other heating mechanisms or to shortcomings in the models (e.g.,
Draine \& Bertoldi 1999).

From the observational standpoint, a conclusive study of the effects
of metallicity on gas temperature in the Magellanic Clouds must await
the availability of more powerful submillimeter instruments observing
the southern sky such as the Atacama Large Millimeter Array (ALMA),
and the recently deployed single--dish telescopes: the Atacama
Pathfinder Experiment (APEX) and the Atacama Submillimeter Telescope
Experiment (ASTE).

\acknowledgements We wish to thank the AST/RO group, and the anonymous
referee. ADB wishes to thank D. Hollenbach for discussions on the
subject of warm molecular gas, and for suggestions that helped improve
this manuscript. ADB and CLM acknowledge support from NSF grants
AST-0228963 and OPP-0126090 respectively.

\label{summ}

\end{document}